\documentclass{iau} 
\usepackage{graphicx}
\usepackage{amsmath}
\usepackage{amsfonts}
\usepackage{amssymb}
\usepackage{graphicx}
\usepackage{float}
\usepackage{bm}
\usepackage[square]{natbib}
\usepackage{indentfirst}

\def\Rt{R_\mathrm{s}}
\newcommand{\pd}{\mathrm{\partial}}
\def\TT{T Tauri }

\title[Dynamo in T Tauri Stars] %% give here short title %%
{Magnetic field generation in PMS stars with and without radiative core}

\author[Zaire et al.] %% give here short author list %% 
{B. Zaire$^{1, *}$, G. Guerrero$^{1, \dag}$, A. G. Kosovichev$^{2}$, P. K. Smolarkiewicz$^{3}$ 
\& N. R. Landin$^{4}$}
 
\affiliation{$^{1}$Physics Department, Universidade Federal de Minas Gerais \\
 Belo Horizonte, MG 31270-901, Brazil\\
$^{*}$ e-mail: zaire@fisica.ufmg.br,  $^{\dag}$ e-mail: guerrero@fisica.ufmg.br \\ 
$^{2}$New Jersey Institute of Technology \\ 
 Newark, NJ 07103, USA \\ % e-mail: alexander.g.kosovichev@njit.edu \\
$^{3}$European Centre for Medium-Range Weather Forecasts \\
 Reading RG2 9AX, UK \\   % e-mail: smolar@ecmwf.int \\
$^{4}$ Campus UFV Florestal, Universidade Federal de Viçosa \\ 
Florestal, MG 35690-000, Brazil} %\\  e-mail: nlandin@ufv.br}

\sympyear{2016}
\pubyear{2017}
\volume{328}  %% insert here IAU Symposium No.
\jname{Living Around Active Stars}
\editors{D. Nandi, A. Valio \& P. Petit, eds.}
\doi{10.1017/S1743921317003970}
\begin{document}

\maketitle

\begin{abstract}
Recent observations of the magnetic field in pre-main sequence stars suggest that
the magnetic field topology changes as a function of age. The presence of a tachocline could be
an important factor in the development of magnetic field with higher multipolar modes.
In this work we performed MHD simulations using the EULAG-MHD code to
study the magnetic field generation and evolution in models that mimic stars at
two evolutionary stages.
The stratification for both stellar phases was computed by fitting stellar 
structure profiles obtained with the ATON stellar evolution code.
The first stage is at $1.1 \mathrm{Myr}$, when the star is completely 
convective. The second stage is at $14 \mathrm{Myrs}$, when the star is 
partly convective, with a radiative core developed up to $30\%$ of the stellar radius.
In this proceedings we present a preliminary analysis of the resulting mean-flows and magnetic field.
The mean-flow analysis shown that the star rotate almost rigidly on the fully convective phase, 
whereas at the partially convective phase there is differential rotation with conical contours of iso-rotation.
As for the mean magnetic field both simulations show similarities with respect to the field evolution.
However, the topology of the magnetic field is different.

\end{abstract}

\firstsection 
\section{Introduction}
Magnetic field plays an important role in the evolution of pre-main sequence (PMS) stars.
At this phase, solar type stars are called \TT stars.
They have a disk which will dissipate within the first ten millions of years of their life. 
The star-disc interaction is mediated by the action of a large-scale magnetic field, 
which governs the accretion process. Both, the disk and the magnetic field strength are key 
factors to understand the angular momentum evolution of these objects \citep{GB13}.

Recently \citet{DJGP07,DJGP08,DSBJ10,DBWG11a,DGAH12} derived 
profiles of the magnetic field for a small sample of \TT stars. 
The differences in those maps suggest that a dynamo mechanism operates in the stellar 
interior generating and sustaining a large-scale magnetic field. 
For the observed \TT stars, \cite{GDMHNHJ12} found a relation
between the large-scale magnetic field and the position in the H-R diagram.
They found that the magnetic field gains complexity with the evolution of the star.
For instance, when the star is completely convective, the magnetic field is 
mainly dipolar. After the development of the radiative core the dipolar 
component looses power compared to high order components of the mutipole expansion.

According to the mean-field theory the dynamo depends on the 
large scale motions, differential rotation and meridional circulation, and
on turbulent convection.  All these properties vary with the age of the star. 
For example, the results of \cite{VGJD14} indicate that for \TT
stars the period of rotation decreases with age.
Unfortunately, apart from very recent observations \citep[e.g.,][]{DCMCW00, DSBJ10}, 
little is known about the differential rotation of these objects. 
As for the meridional circulation the only hint we have is provided by global numerical
simulations \cite[e.g, ][]{GSKM13b, G12}.  Note, however, that these
models correspond to main sequence stars.  Finally, for the convective motions we 
rely on the results of stellar structure and evolution models based on the mixing 
length theory (MLT).

Previous numerical results of rotating turbulent convection in fully convective models 
have shown that as the density stratification progressively increases, it weakens the 
dipole-dominant magnetic field topology \citep{G12}. As an illustration, \cite{B08} 
carried out a highly-stratified simulation for a M dwarf star obtaining a weak dipolar field.
More recently, \cite{RUJTAKS15} carried out a numerical simulation obtaining 
a consistent magnitude and morphology, with a dipolar-dominant surface magnetic field. 
However, as pointed by \cite{BB11}, this picture is different for the stratification of \TT 
stars once the gravitational contraction is still present and the nuclear reactions are absent. 
\citeauthor{BB11} simulated the interior of BP Tau star after the development of a radiative 
core covering the inner $14\%$ of the stellar radius.
Although their results provided important insights on the convective motions, the amplitude of 
the large-scale magnetic field was weaker than expected by the observations. 

Here we perform 3D numerical simulations for the target star BP Tau in two different phases 
of evolution. The first one corresponds to the fully convective phase, corresponding to BP 
Tau today. 
The second one corresponds to the same star once its radiative core has developed until $30\%$ 
of the stellar radius. 
Our main goal is to explore the generation of large-scale flows and magnetic field
in this type of stars and study the role of the tachocline in the dynamo mechanism. 
Our results will allow to verify through numerical experiments the hypothesis presented
in \cite{GDMHNHJ12} regarding the topological differences between the stellar magnetic 
field of stars at different ages. Besides, this study will allow to establish a better
connection between the models and the observations for this kind of objects. 

\section{Numerical model}

We adopt a full spherical shell, $0\le \phi \le 2\pi$, $0\le \theta \le \pi$.
In the radial direction the bottom boundary is located at $r_\mathrm{b} = 0.10\Rt$ and 
the upper boundary is at $r_\mathrm{t} = 0.95 \Rt$, where $\Rt$ is the stellar radius.
Similarly to \cite{GSDKM16a} we solve the set of anelastic MHD equations:

\begin{equation}                                                                                               
{ \nabla}\cdot(\rho_{\mathrm{s}}\bm u)=0, \label{equ:cont}
\end{equation}
\begin{equation}
	\frac{\mathrm{D} \bm u}{\mathrm{D}t}+ 2{\bm \Omega} \times {\bm u} =  
    -{\bm \nabla}\left(\frac{p'}{\rho_{\mathrm{s}}}\right) + {\bf g}\frac{\Theta'}
     {\Theta_{\mathrm{s}}} + \frac{1}{\mu_0 \rho_{\mathrm{s}}}({\bm B} \cdot \nabla) {\bm B} \;, \label{equ:mom} 
\end{equation}
\begin{equation}
	\frac{\mathrm{D} \Theta'}{\mathrm{D}t} = -{\bm u}\cdot {\bm \nabla}\Theta_{\mathrm{e}} -\frac{\Theta'}{\tau}\;, \label{equ:en} 
\end{equation}
\begin{equation}
 \frac{\mathrm{D} {\bm B}}{\mathrm{D}t} = ({\bf B}\cdot \nabla) {\bm u} - {\bm B}(\nabla \cdot {\bm u})  \;,
 \label{equ:in} 
\end{equation}
\noindent where $\mathrm{D}/\mathrm{D}t = \pd/\pd t + \bm{u} \cdot {\nabla}$ is the total
time derivative, ${\bm u}$ is the velocity field in a rotating 
frame with ${\bm \Omega}=\Omega_0(\cos\theta,-\sin\theta,0)$,
$p'$ is a pressure perturbation variable that accounts for both the gas 
and magnetic pressure, 
${\bm B}$ is the magnetic field, and $\Theta'$ is the potential temperature 
perturbation with respect to an
ambient state $\Theta_{\mathrm{e}}$ \cite[as explained in][]{GSKM13b}.
The term $\Theta'/\tau$ maintains a steady axisymmetric solution of the stellar structure 
against the action of convective turbulent motions by restoring the ambient state with 
potential temperature $\Theta_{\mathrm{e}}$ within a time scale given by $\tau$.
Furthermore, $\rho_{\mathrm{s}}$  and $\Theta_{\mathrm{s}}$ are the density and potential 
temperature of the reference state which is chosen to be isentropic 
(i.e., $\Theta_{\mathrm{s}}={\rm const}$) and in hydrostatic equilibrium;   
${\bm g}$ is the gravity acceleration and $\mu_0$ is the magnetic permeability. 
The potential temperature, $\Theta$, is related to the specific entropy: 
$s=c_p \ln\Theta+{\rm const}$. 

The equations are solved numerically using the EULAG-MHD code
\citep{GSKM13b, GCS10,RCGS11,SC13}, a spin-off of the hydrodynamical model EULAG 
predominantly used in atmospheric and climate research \citep{PSW08}.
The ambient and isentropic states, as well as the gravity acceleration, 
have been computed by fitting the stellar structure profiles 
obtained with the ATON stellar evolution code \citep{LVDM06} to simple hydrostatic polytropic models.
For the velocity field we use impermeable, stress-free conditions at the top and bottom 
surfaces of the shell; whereas the magnetic field is assumed to be radial at these 
boundaries. Finally, for the thermal boundary condition we consider zero divergence 
of the convective flux at the bottom and zero flux at the top surface.

\section{Results}
\subsection{Convective structures}

Figure \ref{fig:ur} shows instantaneous snapshots of the radial velocity, $u_r$, for 
models $\mathrm{TT}01$ $(\bf{a})$ and $\mathrm{TT}14$ $(\bf{b})$. The three upper panels in each side
represent the vertical velocity in the Mollweide projection at $30\%$, $60\%$  and $90\%$ of the stellar 
radius, respectively. The bottom panels shows the same component of the velocity in 
a longitudinal plane at the stellar equator.
For model $\mathrm{TT}01$, at the top of the domain 
small scales of convection can be observed 
at hight latitudes, however, the convective patters are less evident
at low-latitudes. 
In the deeper stellar interior the convective patterns become columnar, exhibiting 
the so-called ``banana cells". For the partially convective star we can identify, as expected, the 
existence of the radiative core in the longitudinal cut. Since this layer is stably stratified, the 
vertical motions only slightly overshoot in this region. An important feature observed is the development 
of large vertical upflows and downflows at equatorial latitudes in the upper part of the domain.
These motions contribute to the mean profile of the meridional circulation, as it will be presented below. 
The banana cells are also evident in this models at $r=0.35 r_s$ and $r=0.6 r_s$. 

\subsection{Large-scale flows}

\begin{figure}[h!]
\begin{minipage}{.5\textwidth}
$(\bf{a})$
\includegraphics[width=5.5cm]{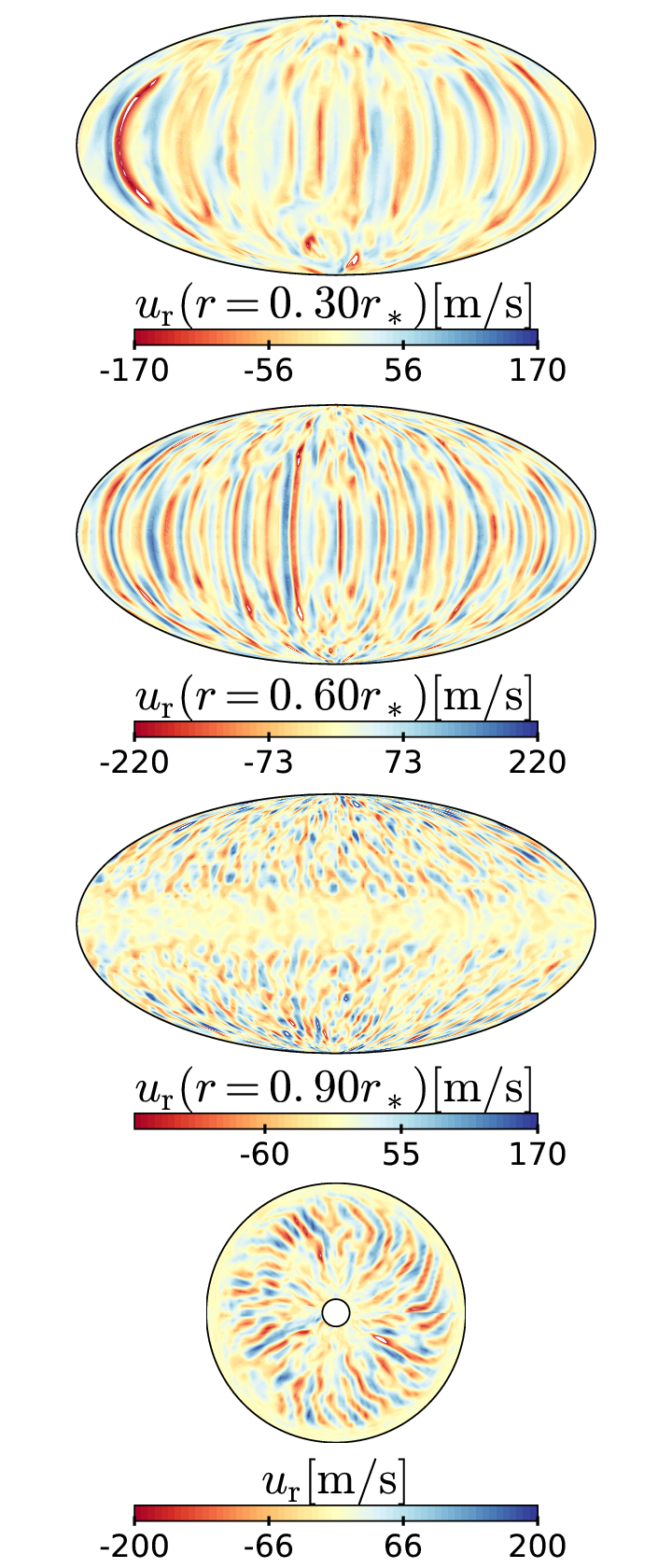} 
\end{minipage}%
\begin{minipage}{.5\textwidth}
$(\bf{b})$
\includegraphics[width=5.5cm]{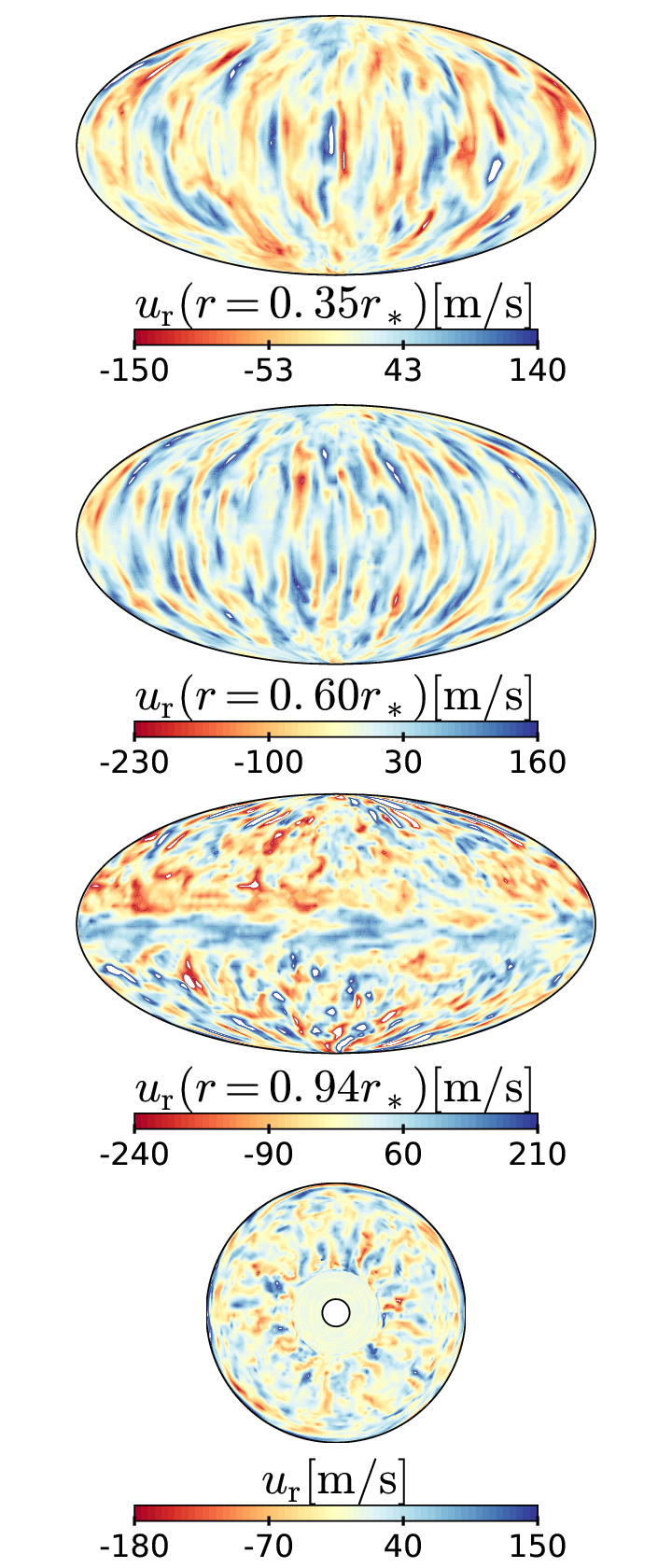}
\end{minipage}
\centering
\caption{ Panels $(\bf{a})$ and $(\bf{b})$ shows instantaneous snapshots of the radial velocity in 
the MHD simulations $\mathrm{TT}01$ and $\mathrm{TT}14$, respectively. 
The three upper panels in each side are Mollweide projections of $u_\mathrm{r}$ at $30\%$, $60\%$ and $90\%$ of the stellar 
radius, respectively. The bottom panel is a longitudinal, $r-\phi$, cut at the stellar equator.\label{fig:ur}}
\end{figure}

The Figure \ref{fig:ur} shows the final moments of the simulations. 
Representing the vertical velocity, they show that the flow patterns 
are distributed uniformly around the rotation axis. 
Therefore, the models are suitable for mean-field analysis. We are interested specially in the large-scale 
flows and magnetic fields (see next section). 

\begin{figure}[htp]
$(\bf{a})$
\begin{minipage}{.29\textwidth}
\includegraphics[width=4cm]{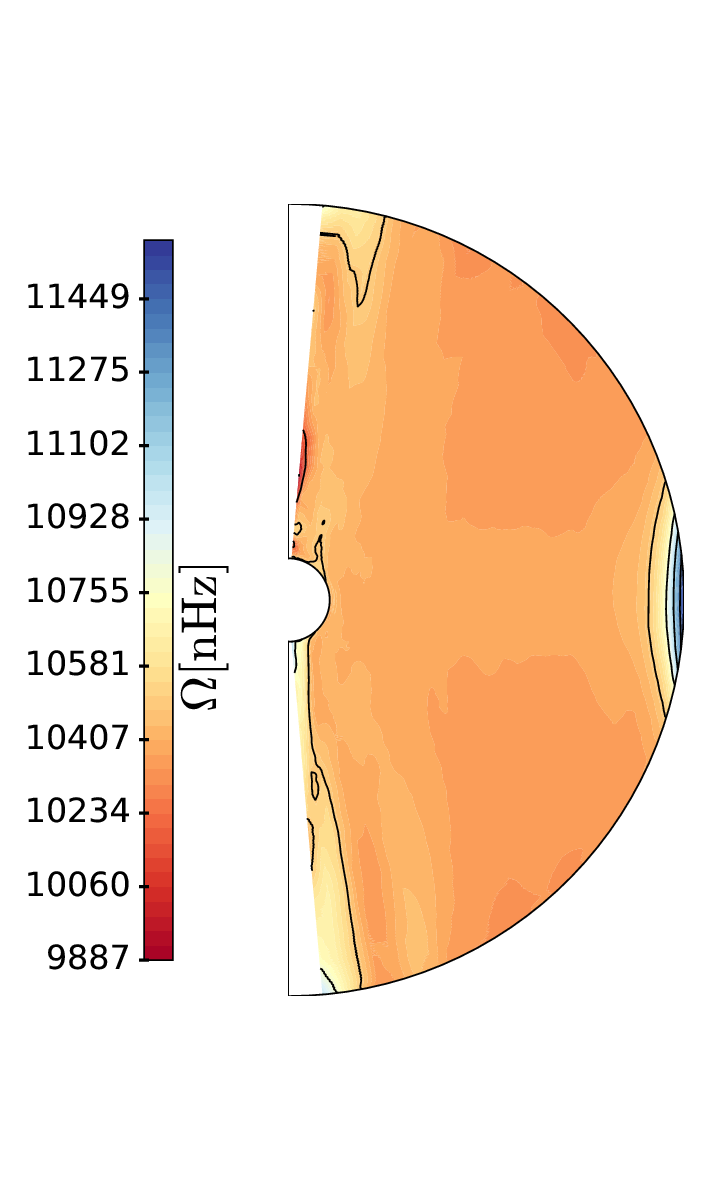}
\end{minipage}%
$(\bf{b})$
\begin{minipage}{.29\textwidth}
\includegraphics[width=4cm]{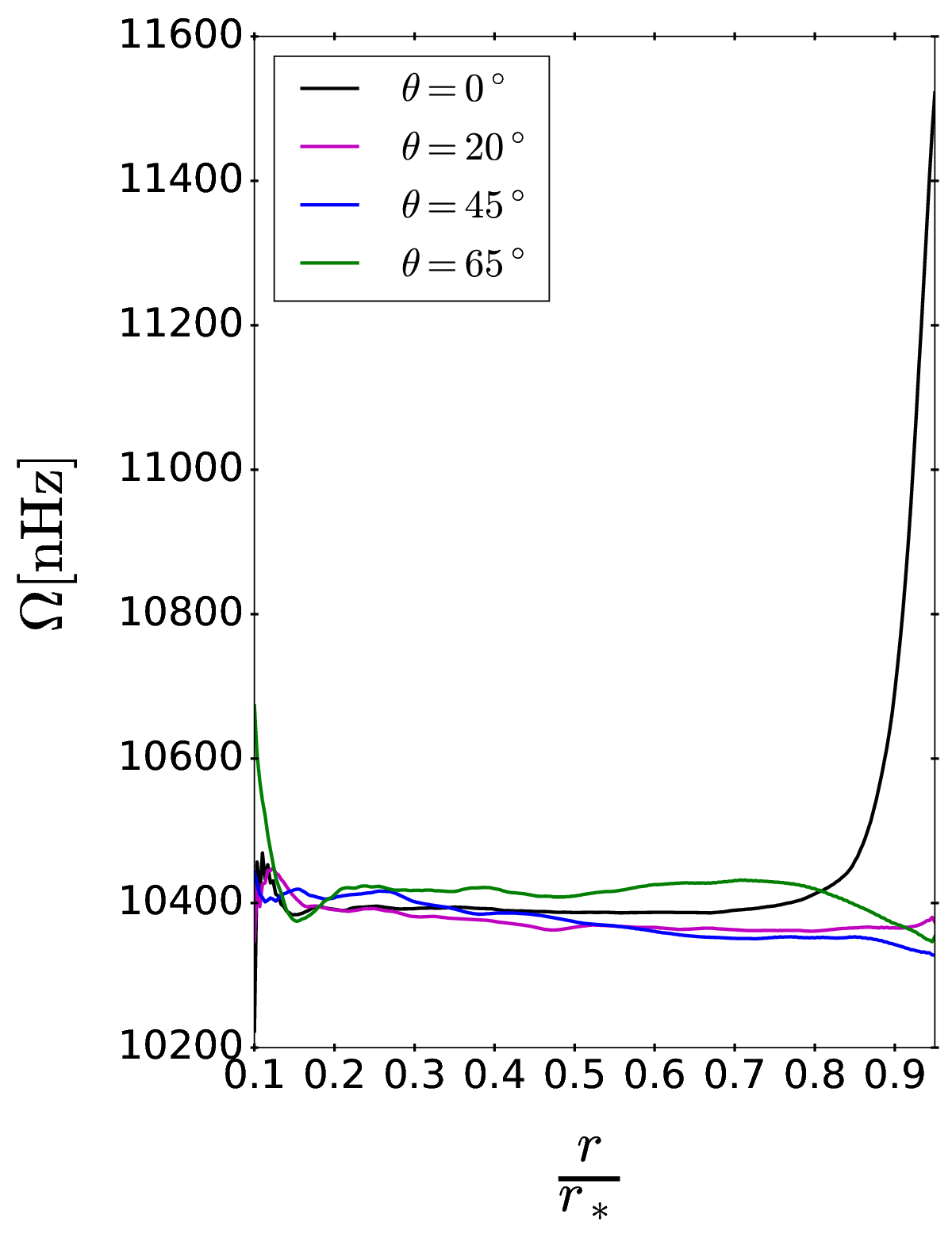}
\end{minipage}%
$(\bf{c})$
\begin{minipage}{.29\textwidth}
\includegraphics[width=4cm]{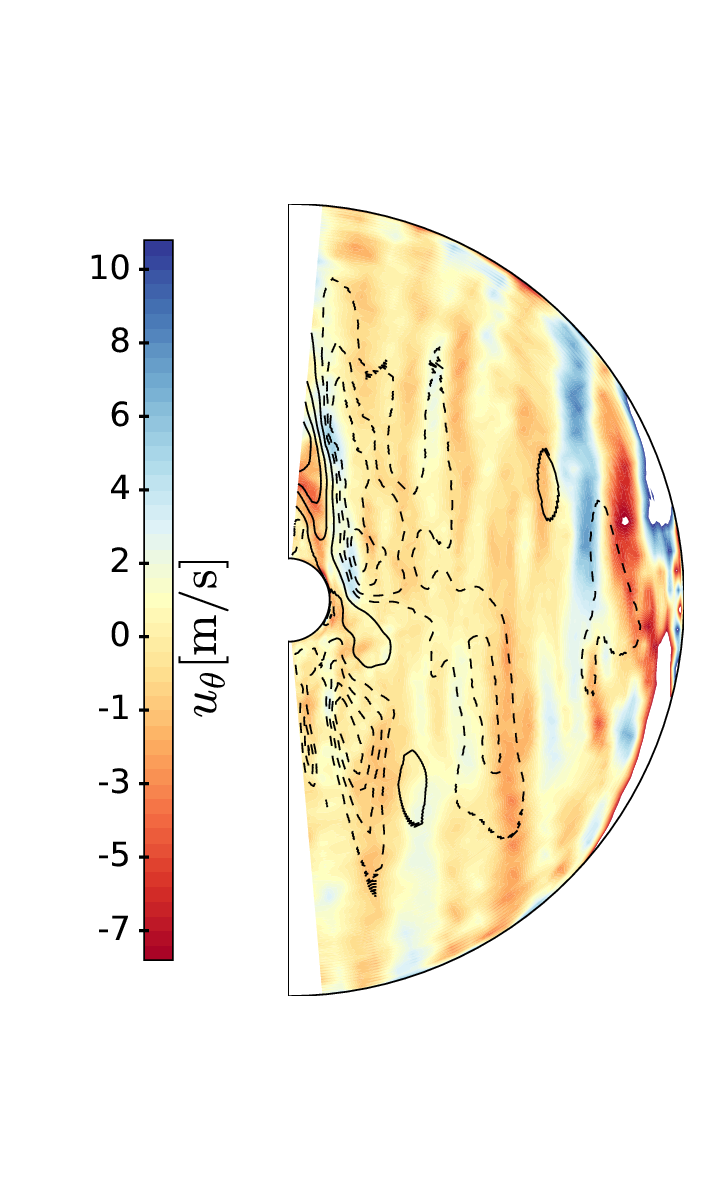}
\end{minipage}
\centering
\caption{Panels $(\bf{a})$  and $(\bf{b})$ present the angular velocity for model $\mathrm{TT}01$.  
Panel $(\bf{a})$ shows the differential rotation profile in the meridional plane, and panel $(\bf{b})$ 
exhibits the radial distribution of the angular velocity at different latitudes. Panel $(\bf{c})$ 
presents the meridional circulation profile in the meridional plane.
The colored contours show $\overline{u}_{\theta}$. 
The contour lines show the stream function $\Psi$, computed from $\rho \bf{u} = \nabla \times \Psi$.  
Solid (dashed) lines correspond to clockwise (counterclockwise) circulation. 
These profiles correspond to azimuthal and temporal averages considered at the final stages 
of the simulation. \label{fig:omega01MHD}}
\end{figure}

\begin{figure}[htp]
$(\bf{a})$
\begin{minipage}{.29\textwidth}
\includegraphics[width=4cm]{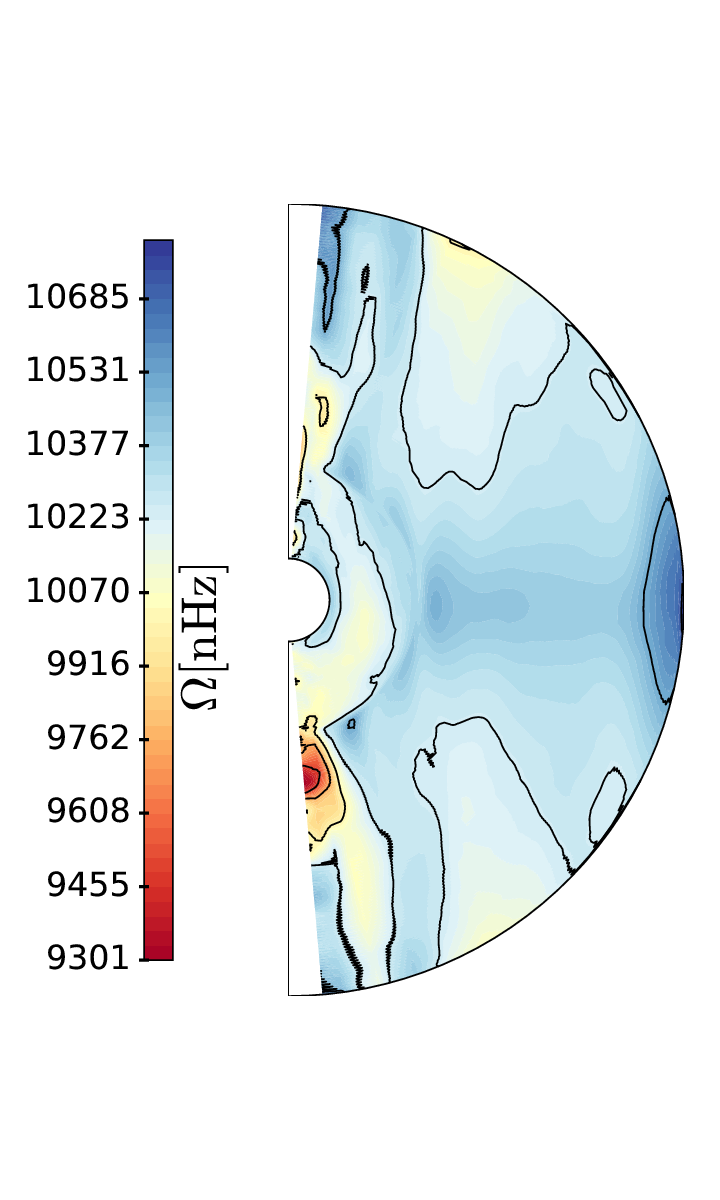}
\end{minipage}%
$(\bf{b})$
\begin{minipage}{.29\textwidth}
\includegraphics[width=4cm]{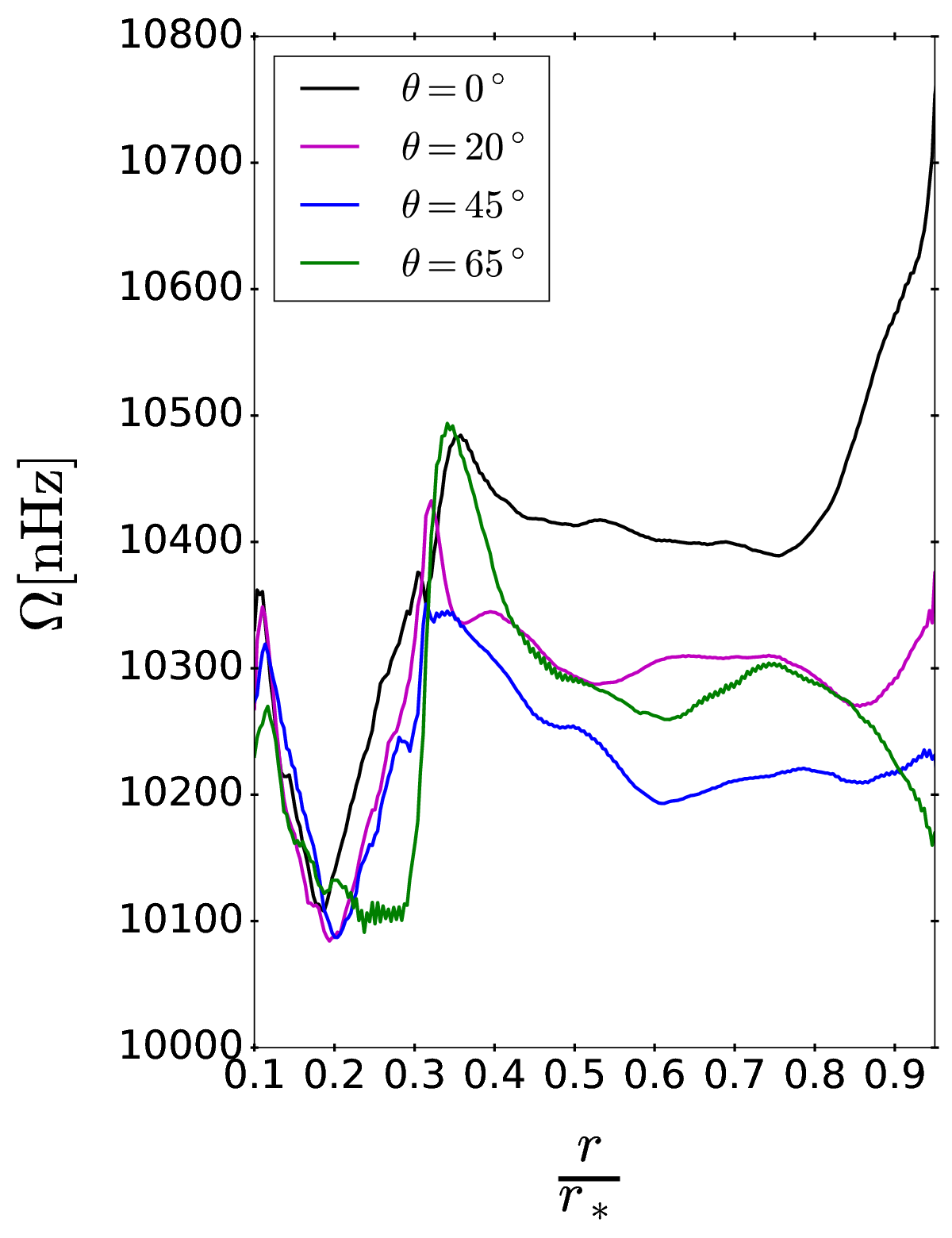}
\end{minipage}%
$(\bf{c})$
\begin{minipage}{.29\textwidth}
\includegraphics[width=4cm]{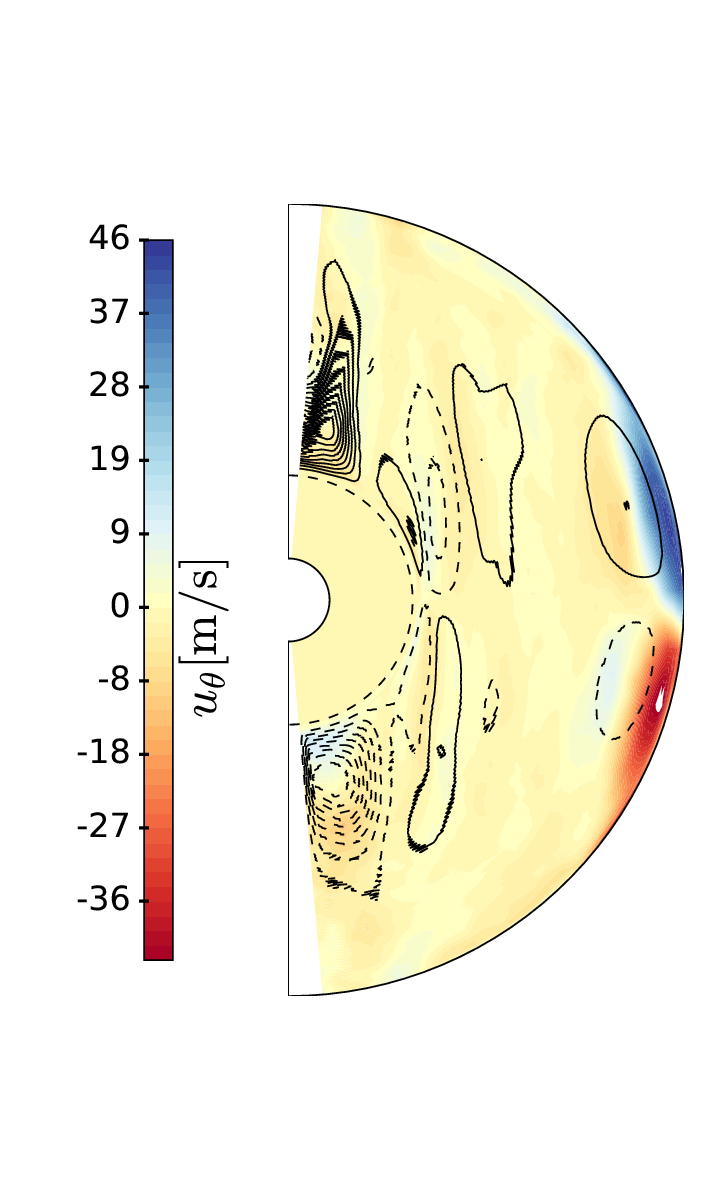}
\end{minipage}
\centering
\caption{The same as in Figure \ref{fig:omega01MHD}, but for model $\mathrm{TT}14$.\label{fig:omega14MHD}}
\end{figure}

Figures \ref{fig:omega01MHD} and \ref{fig:omega14MHD} present
the angular velocity and the meridional circulation for the models 
$\mathrm{TT}01$ and $\mathrm{TT}14$, respectively. It is possible to observe 
that for model $\mathrm{TT}01$ the star rotates almost uniformly, with the exception 
of a prominent shear region in the upper part of the domain at lower latitudes
(Figure \ref{fig:omega01MHD} $\bf{a}$). 
The model $\mathrm{TT}14$ exhibits a conical profile of differential rotation and it is 
possible to notice the presence of the tachocline, although it is not sharply defined 
(Figure \ref{fig:omega14MHD} $\bf{a}$ and $\bf{b}$). 
Both models develop solar-like rotation, which means that the Coriolis 
force dominates over the buoyant convective motions. 
As for the meridional circulation, in both models we observe the existence of circulation cells.
These cells are better structured in the model $\mathrm{TT}14$ than in the model $\mathrm{TT}01$
(compare Figures \ref{fig:omega01MHD} $\bf{c}$ and with \ref{fig:omega14MHD} $\bf{c}$). 
Furthermore, in model $\mathrm{TT}14$ the circulation cells are clearly
symmetric across the equator.

\subsection{Mean-field magnetic fields}

\begin{figure}[htp]
\includegraphics[width=13cm]{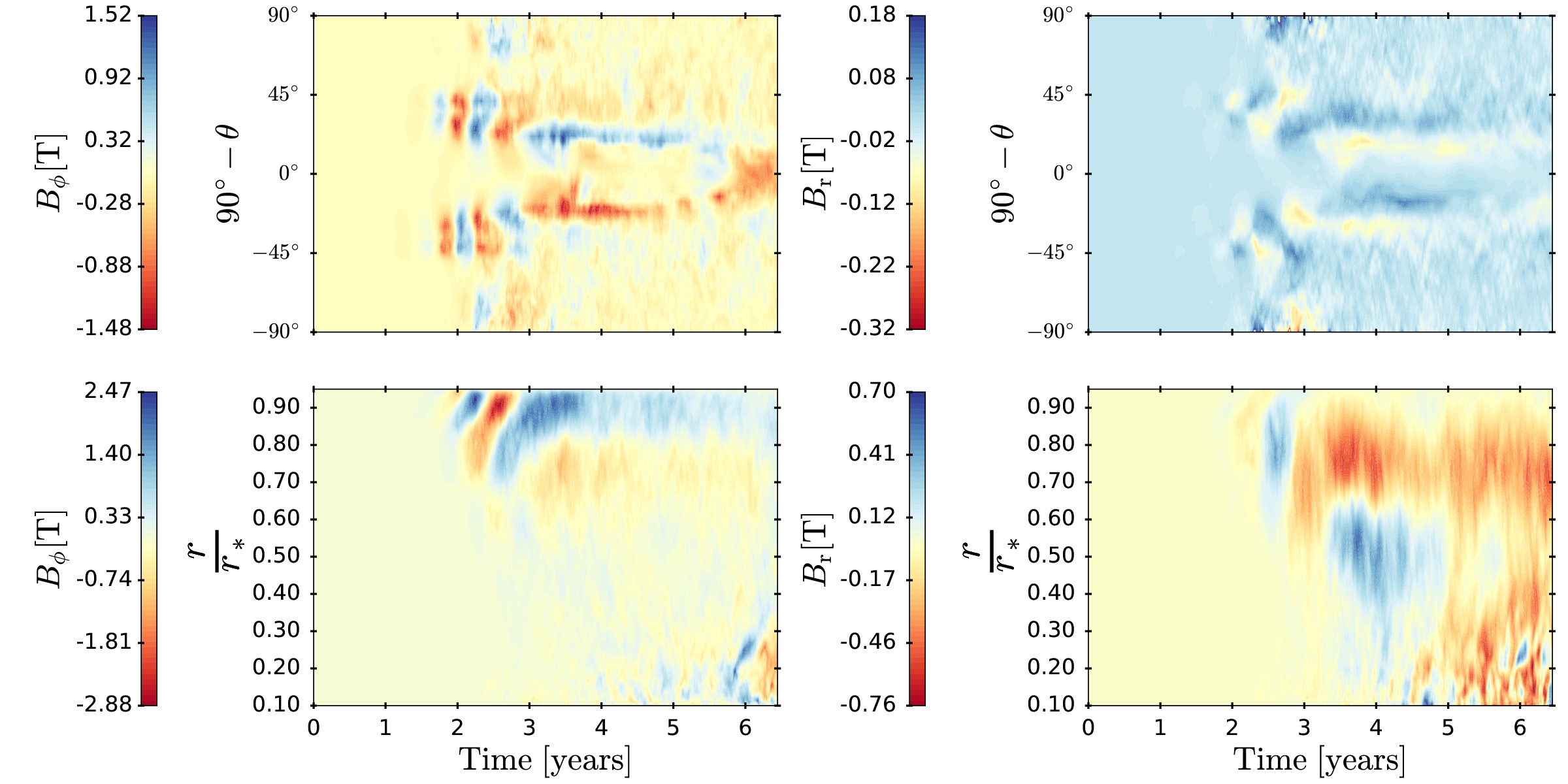}
\centering
\caption{Butterfly diagrams for the fully convective model, $\mathrm{TT}01$. The upper panels show the 
evolution of the toroidal (left) and radial (right) field components in time and latitude at $r = 0.93 \Rt$. 
The lower panels show the time-radius evolution for these quantities at $20^o$ degrees latitude.  \label{fig:mag01}} 
\end{figure}

\begin{figure}[htp]
\includegraphics[width=13cm]{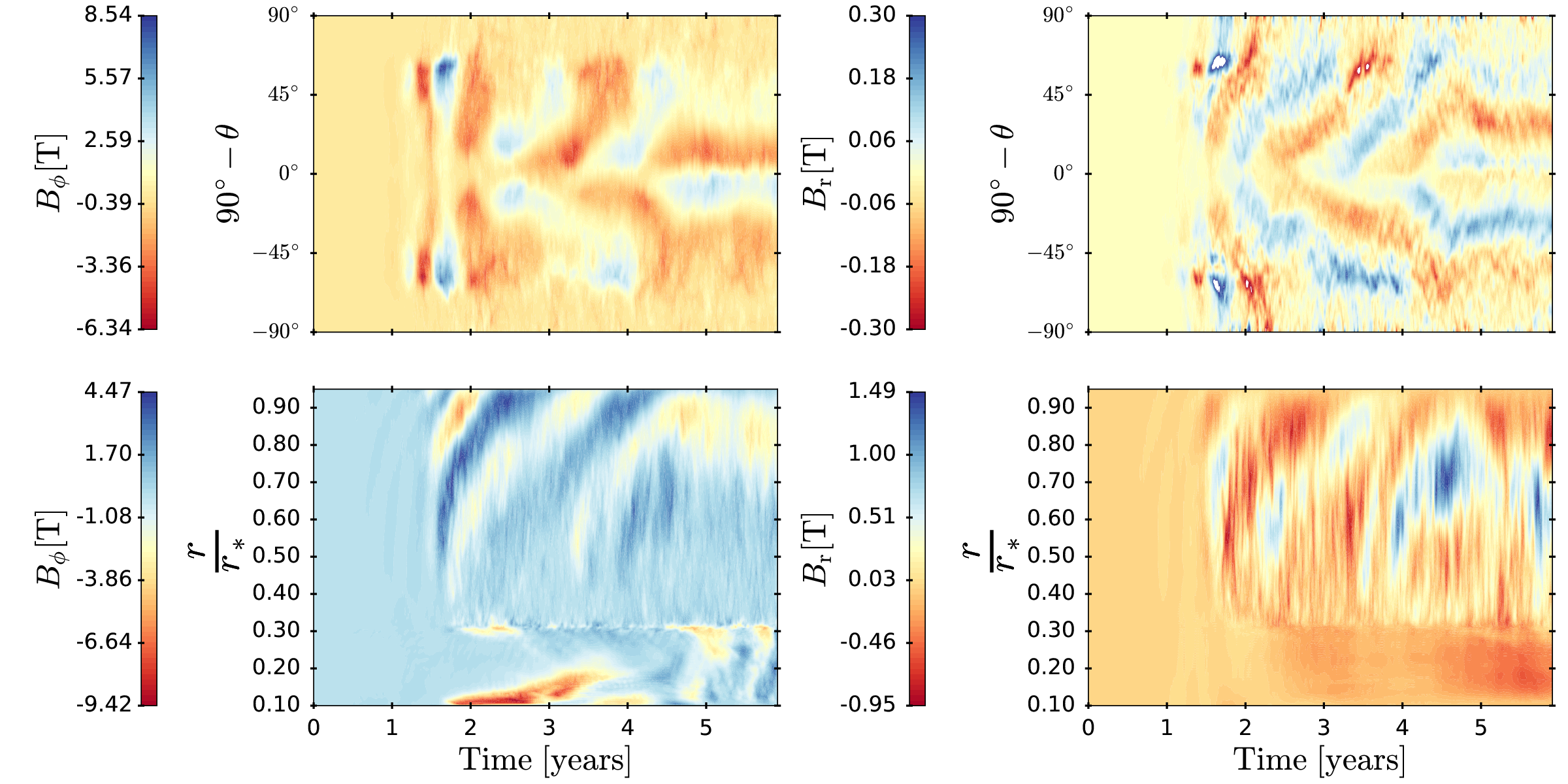}
\centering
\caption{The same as in Figure \ref{fig:mag01} but for the partially convective model, 
$\mathrm{TT}14$. \label{fig:mag14}} 
\end{figure}

Figure \ref{fig:mag01} presents the evolution of the large-scale 
magnetic field obtained in model $\mathrm{TT}01$. The upper (lower) panel is a
time-latitude (time-radius) butterfly diagram for the toroidal (left) and 
radial (right) components of the magnetic field. 
For all component of the magnetic field we notice the existence of a linear phase
during which the magnetic field grows until reaching a magnetic energy of the
same order than the kinetic one. 
The magnetic field then reaches a non-linear stage when it inhibits 
fluid motions via the Lorentz force. At the beginning of this phase
the dynamo is oscillatory and anti-symmetric across the equator,
with dynamo waves that develop near the surface and propagate upwards,
and from mid latitudes towards the equator.
After roughly 3 years  the magnetic field becomes steady for approximately 
more 3 yr. Afterwards a new topological change is observed. 
This indicates that the simulation has not achieved yet its final steady state and
should be ran longer. 
In the non-linear phase of the dynamo, both components of the magnetic field are initially 
generated in the upper part of the domain. As time evolves, the magnetic field develops 
in the entire convective envelope. 
We hypothesize that this  occurs because of the back-reaction of the magnetic field on the flow, 
which gradually establishes new mean-flow profiles (see above) until a final steady state
(probably not yet reached) is obtained. 

The evolution of the mean magnetic field developed in model $\mathrm{TT}14$ 
is presented in Figure \ref{fig:mag14}. 
Similar to the previous case we notice a linear phase
when the magnetic field is amplified until its magnetic energy is comparable 
to the kinetic energy. Following this exponential growing a non-linear phase is observed.  
In this stage, both the toroidal and radial components of the magnetic field are 
initially oscillatory and symmetric across the equator.
The dynamo waves during the first years of the non-linear phase propagate
upwards from the bottom of the convection zone. 
No latitudinal migration of the fields is initially observed, however this migration appears
after about one year of evolution.
The dynamo is then oscillatory and propagates poleward. Both components are initially 
symmetric across the equator but this parity changes after one cycle of evolution. 
A stationary-phase begins at $t \sim 4.5$ years with an anti-symmetric magnetic field. 
The radial large-scale magnetic field corresponds to a quadrupolar topology. 
However, as for case $\mathrm{TT}01$, we still need to run the simulation further to  
be sure that this is the final morphology of the magnetic field. 
It is worth mentioning that in the non-linear phase the magnetic field is generated 
also in the radial shear layer and the radiative core.  
However, the shear does not seem to be sufficient to generate a strong layer of 
magnetic fields.

\section{Conclusions}

Recent observations of the magnetic field in low mass stars have identified different
magnetic morphologies \citep{VGJD14}. The topology of the large-scale magnetic field of these stars 
seems increase in complexity as a function of the stellar age \citep{GDMHNHJ12}.
It has been hypothesized that the appearance of a radiative core changes an 
initially dipolar dominant magnetic field (observed in several fully convective stars),
towards a field with higher multipolar orders.
With the aim to verify this hypothesis, here we studied the generation of magnetic field
in PMS stars through global MHD simulation. Models of stars in two different evolutionary
stages are presented. 
The first one is at $1.1 \mathrm{Myr}$, when the star is completely convective. 
The second stage is at $14 \mathrm{Myrs}$, when the star is partially convective, with a 
radiative core developed until $30\%$ of the stellar radius. 
For this work we performed 3D numerical simulations using the anelastic EULAG-MHD code.
The initial seed magnetic field is random and imposed on the steady state HD solutions. 
The background and environmental stratifications, necessary in our set of anelastic equations,  
are polytropic atmospheres that fit the profiles obtained with the ATON stellar evolution code. 

The results are successful in developing large-scale motions and magnetic fields from
rotating turbulent convection in stars with convective envelopes of different sizes.
For the fully convective model these turbulent motions exhibit small convection
cells at the upper layers at higher latitudes and less pronounced cells at the equator.
For the model with the radiative core the convective structures represent strong downflows in
the convection layer. In the stable layer, the vertical motions only slightly overshoot. 
In deeper layers for both models the convective pattern becomes 
cylindrical with respect to the rotation axis and displays convective "banana cells".
In both models the convective structures are distributed uniformly in the azimuthal direction.
This allows us to perform a mean-field analysis and compute the large scale motions
and magnetic fields.

For the fully convective model the final rotation is almost rigidly, with the existence of shear only in 
the outer layers and at equatorial latitudes.
For the partially convective case the differential rotation is slightly conical with a radial shear 
layer at $r \simeq 0.3 r_s$.
The meridional circulation cells that arise in both simulations seemed to be 
sensitive to the size of the convective domain and to the presence of a large-scale magnetic field. 
For model $\mathrm{TT}01$ the cells are unorganized (this could be an evidence that the model has 
not reached a fully saturated state).
For the partially convective model there is a symmetric multicellular structure across the equator. 
The mean-flows drastically change due to the mean-magnetic field and we expect to reach a saturated 
state when both the mean flows and the mean magnetic field reach equilibrium. 

The magnetic field evolution in the fully convective model, $\mathrm{TT}01$, is 
clearly different from the partially convective model, $\mathrm{TT}14$, which naturally
develops a tachocline. Both models have a linear phase, with exponential growing of 
$\mathrm{\bf{B}}$, as it is characteristic of a mean-field dynamo. Compared to solar dynamo models, for which
non-linear saturation is fast \citep{GSGKM16}, the models presented here have a long non-linear phase 
when both, mean flows and fields mutually adjust. The propagation of dynamo waves differs among the 
models with convection zones of different thickness. At the current state of the simulations, the field 
is steady in both cases, however, their topology is different.  It is still necessary to run the simulations 
further before inferring conclusive results. A more quantitative analysis, regarding angular momentum 
transport as well as studying the source terms of the magnetic field will be presented in forthcoming work
(Zaire et al., 2017, in preparation).

\acknowledgements
This work was partially supported by FAPEMIG grant APQ-01168/14 (BZ and GG) and by NASA grant NNX14AB70G.
We thank the scientific organizers for the wonderful meeting and for the opportunity to present this work. 
BZ acknowledges the IAU, the Physics Department of UFMG for travel support, and 
NASA Ames Heliophysics Summer Program where part of this work was developed. 
PKS is supported by funding received from the European Research Council under the European Union's Seventh 
Framework Programme (FP7/2012/ERC Grant agreement no. 320375). 
The simulations were performed in the NASA cluster Pleiades and in the LNCC cluster SDumont.

\end{document}